\documentclass[a4paper,11pt]{article}
\usepackage{pos}
\usepackage{float}
\usepackage{slashed}

\title{Two-photon contribution to the $K_L\to\mu^+\mu^-$ decay amplitude from a $1/a\approx 1 $ GeV lattice}
\ShortTitle{Two-photon contribution to $K_L\to\mu^+\mu^-$}

\author*[a,1]{En-Hung Chao}
\author*[a]{Ceran Hu}

\affiliation[a]{Physics Department, Columbia University,\\ New York City, New York 10027, USA}

\emailAdd{en-hung.chao@m4x.org}
\emailAdd{ch3704@columbia.edu}

\abstract{We present a lattice-QCD calculation of the long-distance two-photon contribution to the decay amplitude of a long-lived kaon to a pair of muons on a $24^3\times 64$ physical-pion-mass ensemble. In these proceedings, the theoretical foundation of this calculation will be reviewed and preliminary numerical results will be presented for all diagrams with the exception of those that are quark-line disconnected SU(3)-flavor-suppressed. At this stage, the intermediate $\eta$ contribution appears to be the largest source of statistical uncertainty. Our proposed strategies to reduce these statistical errors are discussed in great detail.}

\FullConference{The 41st International Symposium on Lattice Field Theory (LATTICE2024)\\
 28 July - 3 August 2024\\
Liverpool, UK\\}

\note{Supported by the U.S. Department of Energy (DOE) Grant No. DE-SC0011941.}
\note{
This research used resources of the National Energy Research Scientific Computing Center (NERSC), a U.S. Department of Energy Office of Science User Facility located at Lawrence Berkeley National Laboratory, operated under Contract No.~DE-AC02-05CH11231 using NERSC award HEP-ERCAP0023253.
}

\begin{document}
\maketitle

\section{Long-distance part of $K_L\to\mu^+\mu^-$ on the lattice}\label{sec:intro}
Despite its rarity, the $K_L\to\mu^+\mu^-$ decay has been measured very precisely from experiment, leading to a branching ratio of $6.84(11)\times 10^{-9}$~\cite{ParticleDataGroup:2022pth}.
This decay process can be a valuable probe for physics beyond the Standard Model due to its sensitivity to physics at high energy scales. 
However, the long-distance (LD) part due to QED effects in this process is expected to be sizeable compared to the short-distance (SD) part, for which multiple exchanges of $W$ and $Z$-bosons are solely responsible. 
Properly accounting for the interference between the LD and SD parts is thus a critical step in obtaining a complete Standard Model prediction for the decay amplitude.
The current best estimate for the SD part includes the charm-quark effects at next-to-next-to-leading order, which alone amounts to a branching ratio of $0.79(12)\times 10^{-9}$~\cite{Gorbahn:2006bm}.
On the other hand, the most significant contribution of the LD part comes from the two-photon exchange, ie., $K_L\to\gamma^*\gamma^*\to\mu^+\mu^-$~\cite{Martin:1970ai}.
The contribution of this process -- henceforth referred to as LD2$\gamma$ -- to the absorptive part of the amplitude amounts to a branching ratio of $6.59(5)\times 10^{-9}$ from optical theorem~\cite{Cirigliano:2011ny}.
Several estimates for the more challenging dispersive part from phenomenology are available, but the relative sign between the LD and the SD parts cannot be determined easily within the considered frameworks (see Ref.~\cite{Chao:2024vvl} and references therein) and a first-principles, lattice QCD calculation of both the sign and magnitude of this important LD quantity is needed.

A series of previous papers discusses the lattice calculation of this LD2$\gamma$ process and the related lattice calculations of $\pi^0\to e^+e^-$ and $K_L\to\gamma\gamma$ decay~\cite{Christ:2020bzb, Christ:2020dae, Christ:2022rho, Zhao:2022pbs}.
A recent paper~\cite{Chao:2024vvl} presents a detailed concrete proposal whose implementation is treated here and in two earlier proceedings~\cite{Chao:2023cxp, Chao:2024cnu}.
We will give a brief overview of the formalism and its limitations.
Here, the QED effects are treated in the continuum and infinite-volume (QED${}_\infty$) as a coordinate-space integration kernel.
A difficulty in computing the LD2$\gamma$ decay amplitude on the lattice arises as a direct Wick-rotation of the Minkowski-space decay amplitude to its counterpart in Euclidean space is hindered by the $\pi^0$ and $\pi^+\pi^-\gamma$ intermediate states which are less energetic than the initial kaon.
Nonetheless, these being the only intermediate states which lead to unphysical, exponentially-growing contributions under a direct Wick-rotation, one can attempt to identify those contributions in an infrared-regulated theory and extract the physical information. 
The case for the $\pi^0$ is simpler as it is created from the initial kaon through the weak Hamiltonian at the earliest time, and hence must carry vanishing spatial momentum as a consequence of momentum conservation.
The physical contribution of the $\pi^0$ intermediate state can be reconstructed in time-ordered perturbation theory with hadronic matrix elements involving a pion at rest calculated on the lattice.
On the other hand, the case for the $\pi^+\pi^-\gamma$ intermediate states is significantly more complicated: with the photon included in the infinite-volume electromagnetic kernel, the number of hadronic two-pion states leading to unphysical contribution to the decay amplitude grows as we increase the physical size of the lattice, keeping periodic boundary conditions in the spatial directions. 
While, the discrete $\pi^+\pi^-\gamma$ intermediate states which contribute exponentially growing terms could be separately computed and removed, for lattice volumes that are typically available of spatial extent 4-6 fm, all such discrete states are more energetic than the kaon and need not be considered.  

\begin{figure}[h!]
\centering
\includegraphics[scale=0.38]{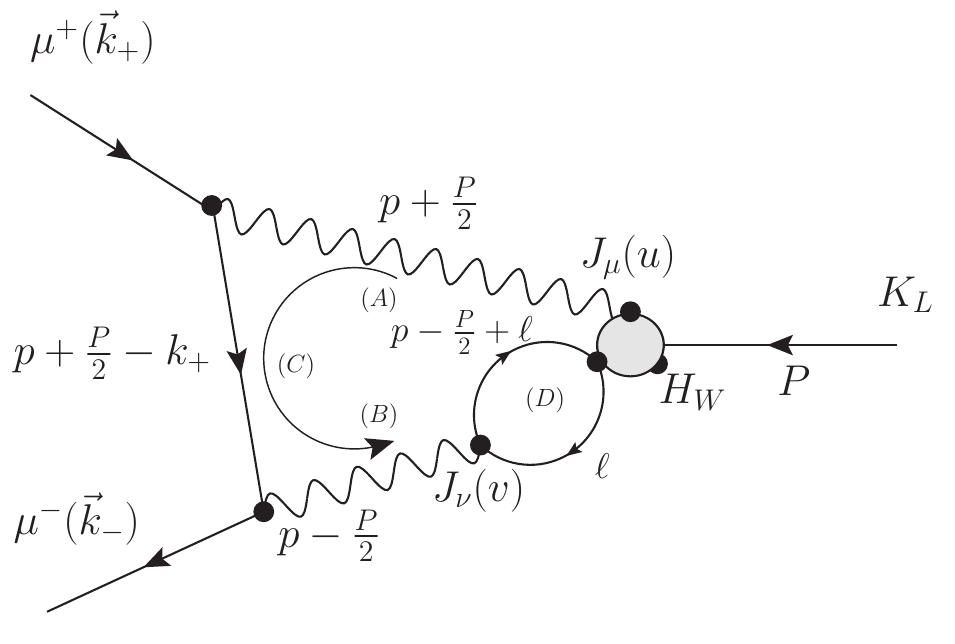}
\caption{Two-pion intermediate state contribution to the LD2$\gamma$ decay amplitude, with possible form factors.}
\label{fig:ppg}
\end{figure}

\begin{figure}[h!]
\centering
\begin{minipage}{0.45\textwidth}
\includegraphics[scale=0.3]{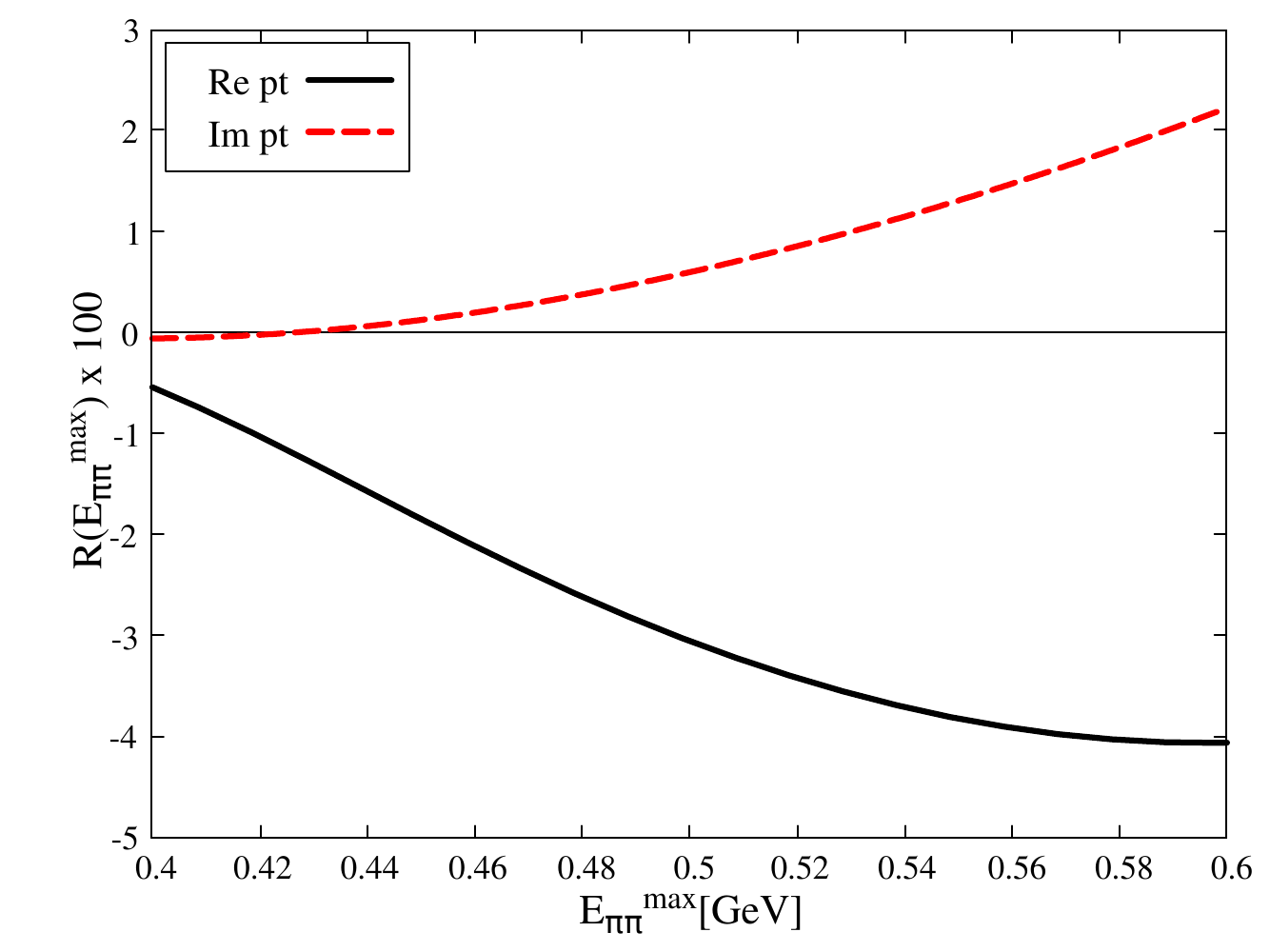}
\end{minipage}
\begin{minipage}{0.45\textwidth}
\includegraphics[scale=0.3]{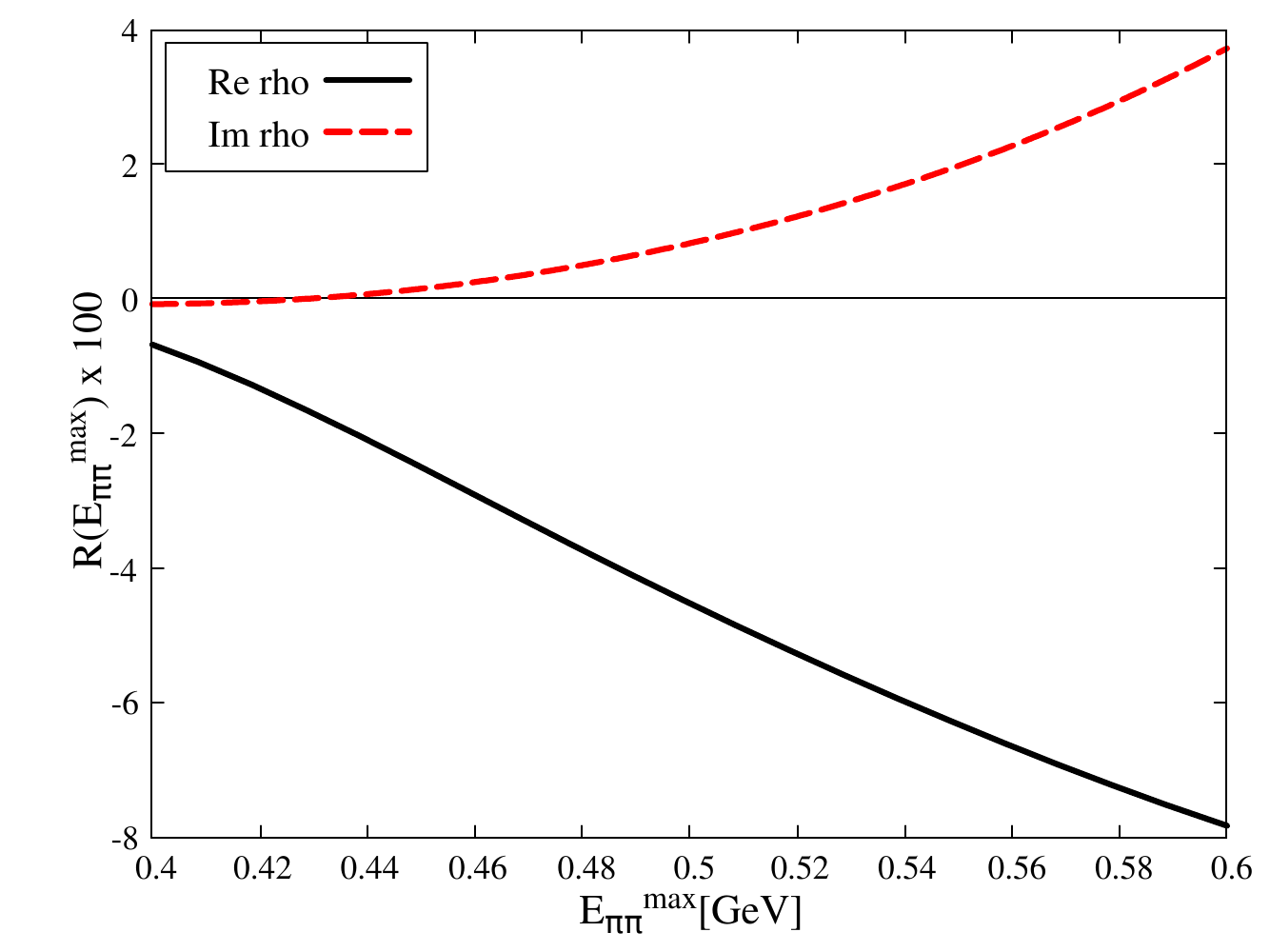}
\end{minipage}
\caption{Real and imaginary parts of the partially-integrated amplitudes normalized to the
amplitude deduced from experiment as a function of the cut-off in the $\pi\pi$ energy in the rest frame
of the kaon, with the $\gamma\pi\pi$ form factor parametrized by a point-like vertex (left) and the Gounaris-Sakurai model (right).}\label{fig:pipig-ratios}
\end{figure}

Concretely, on lattices whose sizes are in the quoted range, the LD2$\gamma$ decay amplitude can be calculated via the following infrared-regulated decomposition: 
\begin{equation}\label{eq:atot}
\mathcal{A} = \mathcal{A}^{\rm I} + \mathcal{A}^{\rm II}\,,
\end{equation}
where 
\begin{equation}\label{eq:a1}
\begin{split}
    \mathcal{A}^{\mathrm{I}} = \int_{-{\delta_{\min}}}^{\delta_{\max}} dv_0\int_V & d^3\textbf{v}
  \int_{v_0}^{R_{\max}+v_0} du_0\int_V d^3\textbf{u}\; e^{M_K(u_0+v_0)/2}\\ 
  &\times L_{\mu\nu}(u-v) \langle T\left\{ J_\mu(u)J_\nu(v) \mathcal{H}_{\rm W}(0) K_L(t_i)\right\} \rangle'\,,
\end{split}
\end{equation}
and
\begin{equation}\label{eq:a2}
\begin{split}
    \mathcal{A}^{\mathrm{II}} = -\sum_n\int_V & d^3\textbf{v}
  \int_{0}^{R_{\max}} dw_0\int_V d^3\textbf{u}\; \left[\frac{e^{M_K w_0/2}}{M_K-E_n}\right]
  \\ & \times L_{\mu\nu}(\textbf{u}-\textbf{v},w_0) \langle T\left\{ J_\mu(\textbf{u},w_0)J_\nu(\textbf{v},0)\right\}|n\rangle
  \langle n|T\left\{\mathcal{H}_{\rm W}(0) K_L(t_i)\right\} \rangle\,.  
\end{split}
\end{equation}
In the above, $L_{\mu\nu}$ is a QED kernel expanded to O($\alpha_{\rm QED}^2$), $J_\mu$'s are electromagnetic (EM) currents, $\mathcal{H}_{\rm W}$ is a $\Delta S=1$ weak Hamiltonian and $K_L$ is a ground-state kaon interpolator.
The usual lattice normalization of state is chosen, such that $\sum_{\psi,\vec{p}}|\psi(\vec{p})\rangle\langle \psi(\vec{p})|=\mathbf{1}$.
We use the prime symbol in Eq.~\eqref{eq:a1} to indicate that the contributions from light single-particle states $|n\rangle$ are removed and Eq.~\eqref{eq:a2} is the physical contribution of those states which should be added back.
Eq.~\eqref{eq:atot} converges to the physical LD2$\gamma$ decay amplitude in Minkowski space when the IR cut-offs, ie. the finite bounds on the temporal integrals in Eq.~\eqref{eq:a1} and Eq.~\eqref{eq:a2}, are removed, and therefore can be evaluated on the lattice.
In practice, in addition to removing the exponentially growing contribution from the state $n=\pi^0$ in order to have convergent quantities in Eqs.~(\ref{eq:a1},\ref{eq:a2}), one could also apply the same procedure to $n=\eta$ to improve the convergence of the total Eq.~\eqref{eq:atot}, where the $\eta$-intermediate state comes with a slowly-falling exponential factor $(e^{(M_K - m_\eta)\delta_{\rm max}}-1)/(M_K - m_\eta)$ (see Sec.~\ref{sec:eta}).
The authors of Ref.~\cite{Chao:2024vvl} estimated the systematic error due to not including the low-energy part of the $\pi\pi$ spectrum resulting from the kinematic constraint imposed by the finite spatial lattice extents.
This estimate was made using a pointlike $K_L\to\gamma\pi^+\pi^-$ vertex and two models for the $\gamma\to\pi\pi$ vertex (see Fig.~\ref{fig:ppg}): (i) pointlike (scalar QED) (ii) Gounaris-Sakurai parametrization, including the effect from the finite width of the $\rho$-meson resonance.
With a cut-off of $E_{\pi\pi}^{\rm max}=0.6$ GeV in the $\pi\pi$ energy in the rest frame of the kaon, the authors conclude that the missing $\pi\pi$ effect is at most 10\% of the real part of the amplitude extracted from experiment and the optical theorem (see Fig.~\ref{fig:pipig-ratios}).
As the current setup is a mixture of finite-volume QCD and infinite-volume QED, one should expect further O($L^{-n}$) suppressed volume effects due to momentum-non-conservation, which formally will contribute to additional exponentially-growing unphysical contributions that are not present in an infinite-volume analysis. 
The size of the latter is more delicate to estimate but is expected to be much smaller than the conservatively quoted 10\% above.

As far as the currently-achieved precision is concerned, a determination of the LD2$\gamma$ amplitude at the ten-percent level can already allow a significant comparison between the Standard Model prediction and experiment.
In the remainder of these proceedings, we report our preliminary results on the 24ID $N_{\rm f}=2+1$ ensemble from the RBC-UKQCD collaboration, which is a substantial extension of the results presented at Lattice2023~\cite{Chao:2023cxp}, with now the disconnected non-SU(3)${}_{\rm f}$-suppressed topology included and strategies for treating the slowly-decaying $\eta$ intermediate state.
Accurately treating the $\eta$ turns out to be the most significant source of statistical uncertainty in the current calculation (see Sec.~\ref{sec:eta}).

\section{Numerical setup}\label{sec:numerics}
\begin{figure}[h!]
\centering
\includegraphics[scale=0.65]{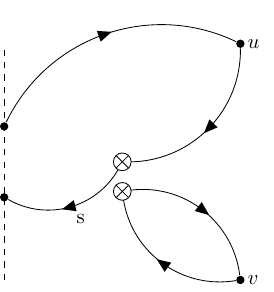}
\hspace{16pt}
\includegraphics[scale=0.65]{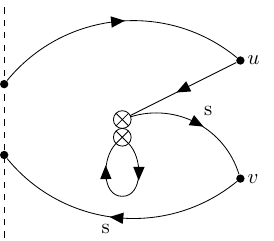}
\hspace{16pt}
\includegraphics[scale=0.8]{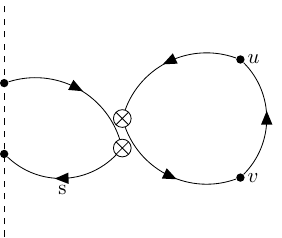}
\\
\includegraphics[scale=0.8]{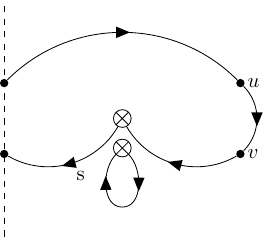}
\hspace{16pt}
\includegraphics[scale=0.8]{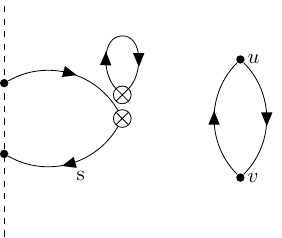}
\caption{The five evaluated non-SU(3)${}_{\rm f}$-suppressed Wick contraction topologies (type-1,-2,-3,-4 and -5 from top to bottom, left to right). The dashed lines indicate the location of the kaon source, the adjacent pair of crosses locate the effective weak Hamiltonian and the dots (labeled $u$ and $v$) represent the electromagnetic currents. Only one representative flavor combination for each topology is displayed.\label{fig:contr}}
\end{figure}

The setup allowing us to properly subtract the unphysical $\pi^0$ intermediate state, and hence to obtain Eq.~\eqref{eq:a1}, is given in our Lattice 2023 proceedings.
As explained in those proceedings, we only consider two of the $\Delta S =1$ operators $Q_1$ and $Q_2$:
\begin{equation}
Q_1 = \left(\bar{s}_a d_a\right)_{V-A}(\bar{u}_b u_b)_{V-A}\,,\quad
Q_2 = \left(\bar{s}_a d_b\right)_{V-A}(\bar{u}_b u_a)_{V-A}\,,
\end{equation}
as the Wilson coefficients for the remaining operators turn out to be much smaller.
Following the non-perturbative renormalization RI/SMOM$(\slashed{q}, \slashed{q})$-scheme as in RBC-UKQCD's recent $K\to\pi\pi$ study~\cite{RBC:2023ynh} using the same gauge field ensemble, the Wilson coefficients are determined to be:
\begin{equation}
C_1 = -0.312 (14)\,,\quad C_2 = 0.718 (14)\,,\quad\textrm{such that}\quad
\mathcal{H}_W =\frac{G_{\rm F}}{\sqrt{2}}V_{us}^* V_{ud}\left( C_1 Q_1 + C_2 Q_2 \right) + \textrm{h.c.}\,,
\end{equation}
where the quoted errors are purely statistical.
As the physical volume of the 24ID ensemble does not allow for exponentially-growing $\pi\pi$-intermediate states, the formalism described in Sec.~\ref{sec:intro} can be applied.
In this work, we only consider the non-SU(3)${}_{\rm f}$-suppressed connected and disconnected diagrams for Eq.~\eqref{eq:a1} as shown in Fig.~\ref{fig:contr}.
We expect the contributions from the omitted diagrams to be small as was the case in the $\pi^0\to e^+e^-$ decay~\cite{Christ:2022rho}.
A high-level description of our implementation strategy for the connected topologies, namely Type-1,-2,-3,-4 topologies shown in Fig.~\ref{fig:contr}, is given in our Lattice 2023 proceedings.
The results presented in these current proceedings are obtained with 512 point-source propagators on 110-120 configurations, separated by 10 Molecular Dynamics trajectories (different statistics for each diagram).
For the Type-5 topology, significantly higher statistics were needed to capture the expected unphysical $\pi^0$ contribution, a contribution which is ultimately cancelled by the $I=0,2$ projection of the quark-connected diagrams.  (See Sec.~\ref{sec:eta}. Unfortunately, it is only the central value, not the noise which is cancelled.) 
In addition to the existing 512 point-source propagators obtained from high precision solves, we increase the statistics by quadrupling the number of point-source propagators sourced at the earliest EM current with lower precision solves, and correct for the bias using the existing more-accurate 512 propagators. 
The variance of the bias is about an order of magnitude smaller than the statistical fluctuation of the data, making this approach efficient~\cite{Chao:2024cnu}. 
We use Coulomb-gauge-fixed wall-source interpolator for the kaon in order to obtain a good overlap with the kaon ground state.
In order to further improve the data quality, we perform an error-weighted average of three to six time-separations, ranging from 6 to 16 lattice units, between the kaon interpolator and the weak Hamiltonian.

To remove the only unphysical, exponentially-growing contribution coming from the $\pi^0$-intermediate state, we compute the $x_0$-summand in the following representation of the amplitude
\begin{equation}\label{eq:a1unsub}
\mathcal{A}(\delta_{\max}) = \sum_{-x_0\leq\delta_{\max}}\sum_{\vec{x},r} L_{\mu\nu}(r)e^{M_Kr_0/2}\langle 0 |\textrm{T}\{ J_\mu(r)J_\nu(0)\mathcal{H}_W(x)\}|K_L\rangle\,,
\end{equation}
where $\delta_{\max}$ is the maximal distance between the earliest EM current and the weak Hamiltonian.
In this representation, the unphysical contribution from the $\pi^0$ reads
\begin{equation}\label{eq:pi0unphys}
\begin{split}
\mathcal{A}_{\pi^0}^{\rm{unphys.}}(\delta_{\max}) = \frac{Ve^{(M_K-m_\pi)\delta_{\max}}}{M_K-m_\pi}&\Theta(\delta_{\max}) \sum_{r}L_{\mu\nu}(r)e^{M_Kr_0/2}
\\ \times &
\langle 0 |\textrm{T}\{J_\mu(r)J_\nu(0)\}|\pi^0(\vec{0})\rangle\langle \pi^0(\vec{0})| \mathcal{H}_W(0)|K_L\rangle\,,
\end{split}
\end{equation}
where $V$ is the spatial volume.
Here we neglect the possibility of the discrete sum introducing O$(a)$ errors which can be avoided by adding some complexity to Eq.~\eqref{eq:pi0unphys}.
The matrix elements required to evaluate the left-hand side of Eq.~\eqref{eq:pi0unphys} can be accurately extracted from three-point functions.
One can obtain Eq.~\eqref{eq:a1} by subtracting Eq.~\eqref{eq:pi0unphys} from Eq.~\eqref{eq:a1unsub} at large-enough $\delta_{\max}$.
On the other hand, Eq.~\eqref{eq:a2} can be obtained by replacing the exponential factor in Eq.~\eqref{eq:pi0unphys} with $(-1)$.

\section{Reconstruction of the physical $\eta$ contribution}\label{sec:eta}
The unphysical contribution from the $\eta$ intermediate state enters in a similar way as the unphysical $\pi^0$ contribution in Eq.~\eqref{eq:pi0unphys} upon appropriate replacements.
The major difficulty in obtaining the physical contribution of the $\eta$ is the slowly-decaying exponential in the unphysical $\eta$ contribution while increasing $\delta_{\rm max}$ due to the closeness between the masses of the kaon and the $\eta$,  which is further enhanced by the energy denominator in Eq.~\eqref{eq:pi0unphys}.
It is challenging to tame the statistical noises which appears in $\eta$-related quantities from the quark-disconnected diagrams. 
Precisely eliminating the unphysical contribution from the $\eta$ and restoring its physical contribution is thus a significant and challenging part of our numerical work.
For this purpose, we have attempted three different methods which should lead to the same physical result. 
As we will see, these three methods give results with similar statistical errors and the agreement of their central values is consistent with controlled systematic errors.

\paragraph{Method 1: direct subtraction}
The most straightforward reconstruction of the physical $\eta$ contribution is to apply the same treatment as used for the unphysical $\pi^0$ as described in Sec.~\ref{sec:numerics}.
One needs to compute Eq.~\eqref{eq:a1} and Eq.~\eqref{eq:a2} with $n=\eta$.

\paragraph{Method 2: modifying $\mathcal{H}_W$ with $c_s(\bar{s}d+\bar{d}s)$}

Adding the operator $\bar{s}d+\bar{d}s$ to the weak Hamiltonian does not alter the on-shell physical decay amplitude that we are computing because this operator is a total divergence of the neutral, strangeness-changing vector current.
To demonstrate the latter, we perform a local infinitesimal flavor rotation between the $d$- and $s$-quarks, ie.
    \begin{eqnarray}\label{eq:rotation}
    &\left(
    \begin{aligned}
        d\\
        s\\
    \end{aligned}
    \right)
    \longrightarrow
    (\mathbf{1}_{2\times 2}+\epsilon\, \mathcal{T})
    \left(
    \begin{aligned}
        d\\
        s\\
    \end{aligned}
    \right), 
    \quad\textrm{where}\quad
\mathcal{T}=
    \left(
    \begin{aligned}
       &0 & 1\\
       -&1 & 0\\
    \end{aligned}
    \right)\,.
    \end{eqnarray}
    This symmetry transformation leads to the Ward identity, valid if the kaon matrix element is extracted without excited state contamination,
    \begin{equation}\label{eq:Ward_M2}
        \left<(m_d-m_s)(\bar{s}d+\bar{d}s)(x){O}_{\mu\nu}(u,v,w)+i\partial_\lambda(\bar{s}\gamma_\lambda d(x)-\bar{d}\gamma_\lambda s(x)){O}_{\mu\nu}(u,v,w)\right> = 0\,,
    \end{equation}
    where ${O}_{\mu\nu}(u,v,w) \equiv J_\mu(u) J_\nu(v) K_{ L}(w)$.
    We can thus add a lower-dimensional operator $c_s(\bar{s}d+\bar{d}s)$ to the weak Hamiltonian without changing the final, on-shell result given in Eq.~\eqref{eq:a1unsub}.
    The coefficients $c_s$ can be tuned to cancel the overlap between different $\Delta S = 1 $ four-quark operators and $\eta$,
    \begin{eqnarray}
   \label{eq:cs_M2}
        c_{s,i} = -\frac{\left<\eta|Q_i|K_L\right>}{\left<\eta|\bar{s}d+\bar{d}s|K_L\right>}\,,\quad \mathcal{H}^\prime_W =\frac{G_{\rm F}}{\sqrt{2}}V_{us}^* V_{ud}\sum\limits_{i=1}^2 C_i \left( Q_i+c_{si}\bar{s}d\right) + \textrm{h.c.}\,.
    \end{eqnarray}
By construction, replacing $\mathcal{H}_W$ by $\mathcal{H}_W^\prime$ leads to the same physical amplitude but $\mathcal{H}_W^\prime$ does not receive contribution from the  $\eta$-intermediate state.

\paragraph{Method 3: modifying $\mathcal{H}_W$ with $c_s(\bar{s}d+\bar{d}s)$ (variant)}
In our setup with $N_{\rm f}=2+1$ quark flavors, the EM currents with physical EM charges can be decomposed unambiguously into two distinct isospin components with $I=0$ and $I=1$, $J_\mu=J_\mu^{I=1}+J_\mu^{I=0}$, where
\begin{eqnarray}\label{eq:J_iso}
J_\mu^{I=1} = \frac{1}{2}\bar{u}\gamma_\mu u -\frac{1}{2}\bar{d}\gamma_\mu d\,,\quad J_\mu^{I=0} = \frac{1}{6}\bar{u}\gamma_\mu u + \frac{1}{6}\bar{d}\gamma_\mu d - \frac{1}{3} \bar{s}\gamma_\mu s\,.
\end{eqnarray}
We can apply the same procedure as introduced for Method 2 but only to the isospin channel which receives contribution from the $\eta$-intermediate state.
Considering the same infinitesimal transformation Eq.~\eqref{eq:rotation} applied to the $I=0$ and $I=2$ part (labeled $I=0,2$ below) of the product of the two EM current $J_\mu(u)J_\nu(v)$, 
\begin{equation}
\left[J_\mu(u) J_\nu(v)\right]_{I=0,2} = 
J_\mu^{I=1}(u)J_\nu^{I=1}(v)+J_\mu^{I=0}(u)J_\nu^{I=0}(v)\,,
\end{equation}
and a different tri-local operator 
\begin{equation}
{O}^\prime_{\mu\nu}(u,v,w)= \left[J_\mu(u) J_\nu(v)\right]_{I=0,2}{K}_{L}(w)\,,
\end{equation}
one can derive the Ward identity with contact terms 
\begin{equation}\label{eq:Ward_M3}
\begin{split}
&\left<(m_d-m_s)(\bar{s}d(x)+\bar{d}s(x)){O}^\prime_{\mu\nu}(u,v,w)+i\partial_\lambda(\bar{s}\gamma_\lambda d(x)-\bar{d}\gamma_\lambda s(x)){O}^\prime_{\mu\nu}(u,v,w)\right> 
\\= &\sum_{i=0,1}\Big\{ \langle \delta^4(x-u)\hat{\mathcal{T}}[J^{I=i}_{\mu}](u)  J_{\nu}^{I=i}(v) \tilde{K}_L(w)
+\delta^4(x-v)J^{I=i}_{\mu}(u)\hat{\mathcal{T}}[J_{\nu}^{I=i}](v) \tilde{K}_L(w)\rangle
\\
&+\langle\delta^4(x-w)J^{I=i}_{\mu}(u)J_{\nu}^{I=i}(v)\hat{\mathcal{T}}[ \tilde{K}_L](w)
\rangle\Big\}\,,
\end{split}
\end{equation}
where $\hat{\mathcal{T}}$ acts on the isospin components in the following way
\begin{equation}
\hat{\mathcal{T}}[J^{I=0}_\mu](u) 
=-\hat{\mathcal{T}}[J^{I=1}_\mu](u) 
=\frac{1}{2} \left(\bar{s}\gamma_\mu d+ \bar{d}\gamma_\mu s\right)(u)\,.
\end{equation}
Note that the last term in Eq.~\eqref{eq:Ward_M3} does not contribute to the amplitude in our setup, as the kaon interpolater will always be put away from the rest.
Compared to Method 2, an advantage of this approach is that the computed quantities are not sensitive to the statistical noise from the intermediate $\pi^0$ and other $I=1$ states.
\\

\begin{figure}[h!]
    \centering
    \includegraphics[scale=0.60]{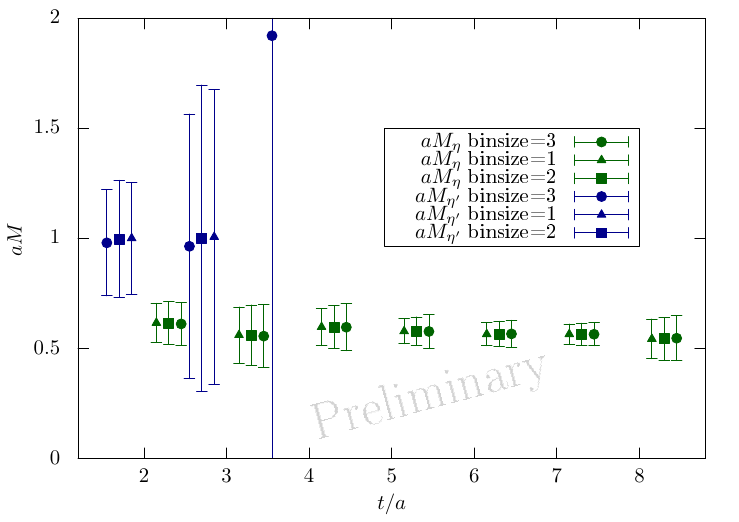}
    \caption{Effective masses of $\eta$ and $\eta^\prime$ from the GEVP on the 24ID ensemble. Here the two times used in the GEVP are $t$ and $t+1$, where $t$ appears on the $x$-axis.}
    \label{fig:GEVP}
\end{figure}

All three methods presented above require computing matrix elements with an external $\eta$ state.
To obtain a good overlap with the $\eta$ ground state, we solve a $2\times 2$ Generalized Eigenvector Problem (GEVP) with the two following Coulomb-gauge-fixed pseudoscalar operators projected to zero spatial momenta:
\begin{equation}\label{eq:GEVP_OPE}
    O_l = \frac{i}{\sqrt{2}}\left(\bar{u}\gamma^5u+\bar{d}\gamma^5d \right)\,,\hspace{1 cm} O_s = i\bar{s}\gamma^5 s\,.
\end{equation}
The preliminary effective-mass plot from solving the GEVP with 519 configurations is presented in Fig.~\ref{fig:GEVP}. 
Note that the early formation of plateaus for $\eta$ and $\eta^\prime$ masses is a consequence of the rather large lattice spacing of the ensemble. 
The effective masses of $\eta$ and $\eta^\prime$ from GEVP read
\begin{eqnarray}\label{eq:GEVP_mass}
am_\eta = 0.569(19)\,, \hspace{1 cm} am_{\eta^\prime} = 0.996(60)\,,
\end{eqnarray}
where the fitting range is $t\in[4,8]$ for $am_\eta$ and $t\in[2,4]$ for $am_{\eta^\prime}$.

To obtain the $\eta$-related matrix elements necessary for all three methods, namely $\langle \eta| Q_{1,2}|K_L\rangle$ and $\langle\eta |\bar{s}d + \bar{d}s| K_L\rangle $, we notice that it is better to take advantage of the small statistical fluctuation when using $O_s$ alone as an interpolator for the $\eta$ ground state\footnote{This observation was made in the Ph.D. thesis of Duo Guo~\cite{Guo:2021bxb}.}. 
Therefore, we perform a two-state fit on the operator-insertion-sink-separation dependence of each relevant three-point function after projecting to the kaon ground state
\begin{equation}\label{eq:2sf}
A(t) = A_\eta e^{-m_\eta t} +A_{\eta^\prime} e^{-m_{\eta^\prime} t}\,,
\end{equation}
to extract the desired matrix elements $A_\eta$ and $A_{\eta^\prime}$.
In Eq.~\eqref{eq:2sf}, the meson masses are taken from the GEVP results Eq.~\eqref{eq:GEVP_mass} with statistical correlations between the two- and three-point functions taken into account by using the jackknife method.

\section{Preliminary results}\label{sec:results}
Before discussing our preliminary results on the 24ID ensemble, we comment on the possible power-law divergence originating from the effective weak Hamiltonian. 
From na\"ive power-counting, one should expect $1/a^2$-divergence from those diagrams with a closed, self-contracted quark loop, ie. those from Type-2, -4 and -5, while taking the continuum limit. 
These divergences enter as coefficients of the dimension-three operator $\overline{s}d+\overline{d}s$. 
While these terms must be removed in an off-shell non-perturbative renormalization scheme~\cite{RBC:2023ynh}, they don't contribute to on-shell amplitudes as discussed above.
However, a useful test of our lattice results is to examine an alternative scheme in which we remove these $1/a^2$ terms directly as follows:
\paragraph{Method 4: modifying $\mathcal{H}_W$ with $d_s(\bar{s}d+\bar{d}s)$ to eliminate $1/a^2$-power divergence}
Starting from Eq.~\eqref{eq:Ward_M2} in Method 2, we consider the modified Hamiltonian
\begin{eqnarray}\label{eq:d_M4}
\mathcal{H}^{\prime\prime}_W =\frac{G_{\rm F}}{\sqrt{2}}V_{us}^* V_{ud}\sum\limits_{i=1}^2 C_i \left( Q_i+d_{s,i}\bar{s}d\right) + \textrm{h.c.}\,,\quad
\textrm{where}\quad
d_{s,i} = -\frac{\left<\pi^0|Q_i|K_L\right>}{\left<\pi^0|\bar{s}d+\bar{d}s|K_L\right>}\,.
\end{eqnarray}
This modified Hamiltonian effectively forbids the $K_L\to \pi^0$ transition and defines a renormalization scheme which is free of $1/a^2$-power divergence~\cite{Bernard:1985wf}.
After this procedure, the unphysical $\eta$-contribution needs be directly subtracted from the modified effective weak Hamiltonian $\mathcal{H}^{\prime\prime}_W$ in the same way as in Method 1. 
\\

\begin{figure}[h!]
    \centering
\begin{minipage}{0.45\textwidth}
    \includegraphics[scale=0.6]{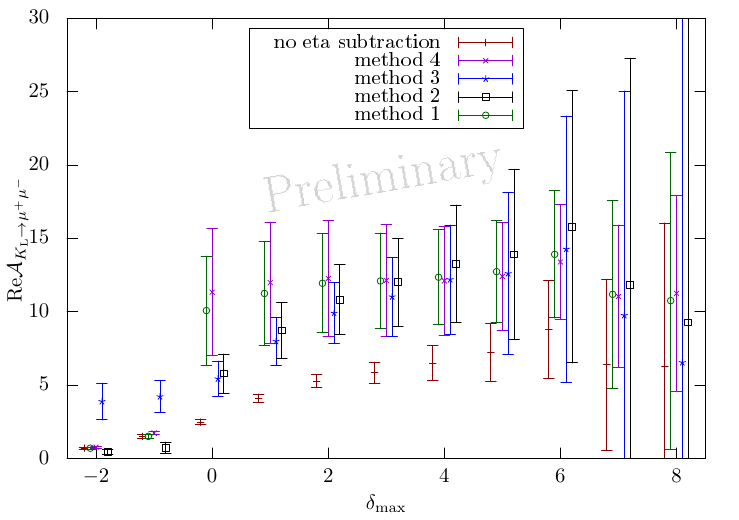}
\end{minipage}
	\hspace{16pt}
\begin{minipage}{0.45\textwidth}
    \includegraphics[scale=0.6]{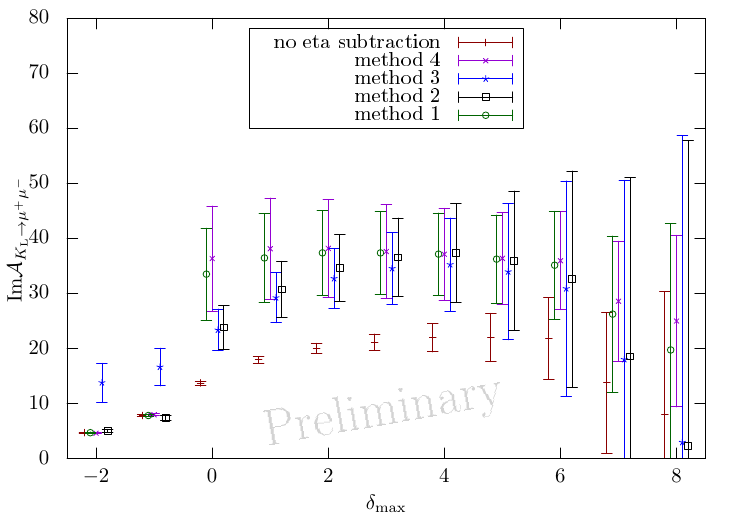}
\end{minipage}
\caption{The real (left) and imaginary (right) parts of the two-photon decay amplitude of $K_L\to\mu^+\mu^-$ on the 24ID ensemble, with $R_{\max} = 7$. ``no eta subtraction'' refers to the result without removing the unphysical $\eta$ contribution.  The $x$-axis is in lattice units and the $y$-axis is in arbitrary units.}
    \label{fig:RealImag}
\end{figure}

The preliminary results of the two-photon decay amplitude of $K_L\to\mu^+\mu^-$ on the 24ID ensemble are presented in Fig.~\ref{fig:RealImag}.
Both plots are composed of results from all four methods sketched so far, as well as the bare results before removing the unphysical $\eta$ contribution. 
The data without removing the unphysical $\eta$ contribution exhibit a long, slowly rising tail possibly visible at large times, where the statistical fluctuations become sizeable.
Such behavior justifies the necessity of the elimination of the unphysical $\eta$ contribution.
The results from all four methods agree within the quoted statistical errors at rather small $\delta_{\max}> 0$ where the data points can be fit to a plateau.
The consistency between Methods 1,2 and 3 is solid evidence of success in dealing with the unphysical $\eta$ contribution. 
On the other hand, the good agreement between Method 4 and the other treatments shows consistency among the different methods for dealing with the divergent $\overline{s}d+\overline{d}s$ term.
The current error budget is dominated by the reconstruction of the physical $\eta$-contribution: by comparing the results from the four methods and the result without removing the unphysical $\eta$ contribution, the size of statistical errors are visibly enhanced for the former. 

\begin{figure}[h!]
\centering
\begin{minipage}{0.45\textwidth}
\includegraphics[scale=0.58]{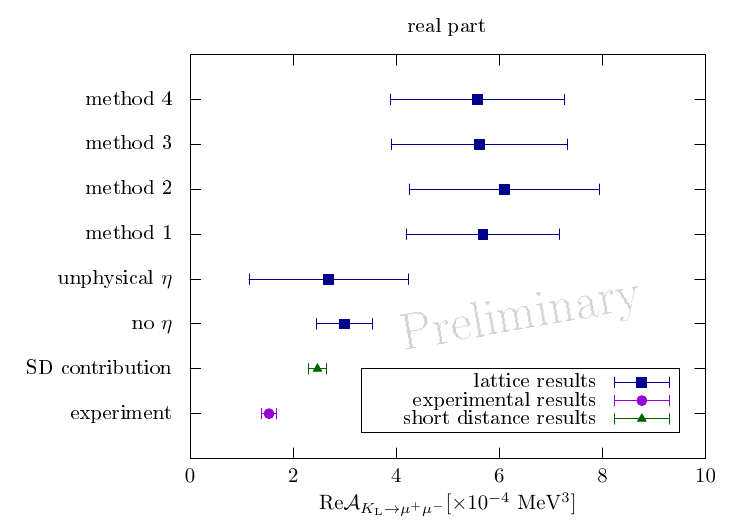}
\end{minipage}
\begin{minipage}{0.45\textwidth}
\includegraphics[scale=0.58]{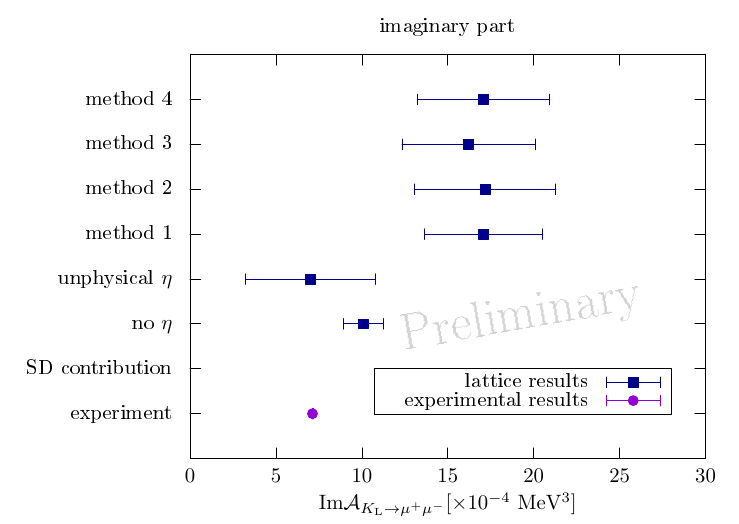}
\end{minipage}
\caption{Comparison between the lattice and experimental results of the real part (left) and imaginary part (right) of the two-photon decay amplitude of $K_L\to\mu^+\mu^-$, alongside the short-distance result. ``no $\eta$'' refers to the result without unphysical-$\eta$ subtraction, ``unphysical $\eta$'' refers to the unphysical-$\eta$ contribution, ``SD contribution'' refers to the short-distance contribution to the real part from the perturbative calculation~\cite{Gorbahn:2006bm}.}\label{fig:exp}
\end{figure}

Using the convention relating the decay amplitude to the decay rate $\Gamma_{K_L\to\mu^+\mu^-}$,
\begin{equation}
    \left|\mathcal{A}_{K_{\rm L}\rightarrow\mu^+\mu^-}\right| = \frac{1}{e^4|V_{ud}V^*_{us}|\frac{G_F}{\sqrt{2}}}\sqrt{\frac{8 \pi M_K \Gamma_{K_{\rm L}\rightarrow\mu^+\mu^-}}{\beta}}\,,\quad
    \beta = \sqrt{1-\frac{4m_\mu^2}{M_K^2}}\,,
\end{equation}
we provide a direct comparison between our lattice calculations for the LD2$\gamma$ decay amplitude, the perturbative SD result and the total decay amplitude extracted from experiment in Fig.~\ref{fig:exp}.
The lattice results are read off from time-slice $\delta_{\max}=4$ with $R_{\max}=7$.  
In Fig.~\ref{fig:exp}, the size of unphysical $\eta$ contribution is also attached, which can be computed from Eq.~\eqref{eq:pi0unphys} with proper replacements. 
We quote our preliminary result from Method 1 which has the smallest absolute error and compare it to the SD and experimental values in Tab.~\ref{tab:res}.
\begin{table}[htbp]
    \centering
    \begin{tabular}{c|c|c}
        & ${\rm Re}\mathcal{A}_{K_{\rm L}\rightarrow\mu^+\mu^-}$ & ${\rm Im}\mathcal{A}_{K_{\rm L}\rightarrow\mu^+\mu^-}$ \\
         \hline
         \hline
         exp.        & 1.53(0.14) & 7.12(0.03)\\
         SD          & 2.47(0.18) & --- \\
         this work (prelim.) & 5.68(1.49) & 17.09(3.43)\\
    \end{tabular}
    \caption{Comparison between theory predictions and experimental values on the decay amplitudes in units of $10^{4}$ MeV${}^3$. The signs here are just convenient choices without physical meanings.}
	\label{tab:res}
\end{table}
For the real-part result, the relative sign between the LD and the SD results being yet to be determined, which will be a main subject of a publication in preparation.
Consequently, a comparison between theory and experiment for $\textrm{Re}\mathcal{A}_{K_L\to\mu^+\mu^-}$ can not yet be made. 
For the imaginary part, we observe a $3\sigma$-discrepancy between the lattice and experimental results at the simulated lattice spacing.
We hope that further studies on finer lattices to understand the scaling behavior of the decay amplitude obtained from our formalism will be helpful in recognizing possible sources of additional systematic error.

\section{Conclusion and outlook}
In these proceedings, we present the work in progress on the lattice determination of the two-photon contribution to the complex $K_L\to \mu^+\mu^-$ decay amplitude. 
The presented numerical work follows the lattice-QCD formalism and the strategy proposed in Ref.~\cite{Chao:2024vvl}. 
Although compromised by the low-energy $\pi\pi\gamma$ intermediate-state contribution, the formalism is expected to give a determination of the decay amplitude with a systematic error arising from such $\pi\pi\gamma$ intermediate states at the 10\% level or below.
One of the difficulties of the formalism is to reconstruct the contribution from the slowly-converging $\eta$-intermediate state.
In this work, we have considered three different methods to remove the unphysical contribution of the $\eta$-intermediate state, which all turn out to be consistent with each other within the quoted statistical error.
Our preliminary result on a $1/a=1.023$ GeV physical pion mass $N_{\rm f} = 2+1$ Domain Wall Fermion ensemble reaches a statistical precision of 25\% for the real part and 20\% for the imaginary part of the LD2$\gamma$ decay amplitude.
The reconstruction of the $\eta$-intermediate state dominates the current statistical error.
A fourth method with a different treatment of the power-law divergent $\overline{s}d+\overline{d}s$ term has also been implemented. 
After the reconstruction of the $\eta$-intermediate state, the shift between this fourth method and the three other is negligible compared to the statistical error, making the treatment of the $\overline{s}d+\overline{d}s$ term less likely to be responsible for the unexplained discrepancy between our lattice result and the experimental value on the imaginary part of the decay amplitude extracted from optical theorem. 
To further investigate this apparent discrepancy on this rather coarse lattice, we have begun simulations on a finer, $1/a=1.73$ GeV, $48^3\times 96$ physical pion mass ensemble. 

\section{Acknowledgement}
We thank our colleagues from the RBC-UKQCD collaboration for useful discussions and substantial technical support.

\end{document}